\documentclass[aps,prl,twocolumn,superscriptaddress,final]{revtex4-2}
\usepackage{amsmath}
\usepackage{amsfonts}
\usepackage{amssymb}
\usepackage{bbm}
\usepackage{graphicx}
\usepackage{xcolor}
\usepackage[colorlinks=true,citecolor=blue]{hyperref}
\usepackage{hypcap}
\usepackage{braket}
\usepackage{bm}
\usepackage{dsfont}
\usepackage{mathtools}
\usepackage[export]{adjustbox} 
\usepackage{verbatim}

\usepackage{csquotes}
\usepackage[backgroundcolor=white,bordercolor=red,linecolor=red,textcolor=red,textsize=footnotesize]{todonotes}

\graphicspath{{.}{./tikz/}}

\DeclareMathOperator{\Tr}{Tr}

\begin{document}
\def\papertitle{Entanglement and private information in many-body thermal states}
\title{\papertitle}
\author{Samuel J. Garratt}
\affiliation{Department of Physics, University of California, Berkeley, California 94720, USA}
\author{Max McGinley}
\affiliation{TCM Group, Cavendish Laboratory, University of Cambridge, Cambridge CB3 0HE, UK}
%\date{\today}
\def\authornames{Samuel J. Garratt and Max McGinley}

\begin{abstract}
We use concepts from quantum cryptography to relate the entanglement in many-body mixed states to standard correlation functions. If a system can be used as a resource for distilling private keys---random classical bits that are shared by spatially separated observers but hidden from an eavesdropper having access to the environment---we can infer that the state of the system is entangled. For thermal states, we derive a simple relation between the information accessible to the eavesdropper and the linear response of the system. This relation allows us to determine which spatial correlations can be used to detect entanglement across wide varieties of physical systems, and provides a new experimental probe of entanglement. We also show that strong symmetries of a density matrix imply the existence of correlations that are always hidden from the environment. This result implies that, although grand canonical ensembles are separable above a finite temperature, canonical ensembles are generically entangled at all finite temperatures.
\end{abstract}

\maketitle

The study of entanglement has provided a unifying description of correlations in many-body quantum ground states \cite{vidal2003entanglement,calabrese2004entanglement,kitaev2006topological,verstraete2006criticality,hastings2007area,chen20101local}. Extending this program of research to finite temperatures promises similar conceptual advances. 
However, unlike pure states, mixed quantum states can exhibit correlations that are entirely classical, and distinguishing these from entanglement is a challenge in macroscopic systems \cite{castelnovo2007entanglement,wolf2008area,frerot2019reconstructing,lu2020structure,kuwahara2022exponential}. In this Letter, we show how correlations between local observables in many-body thermal states can be related to entanglement through the operational task of \emph{quantum key distillation} (QKD)---an important protocol in quantum cryptography \cite{ekert1991quantum}.

\begin{figure}
    \centering
    \includegraphics[scale=1]{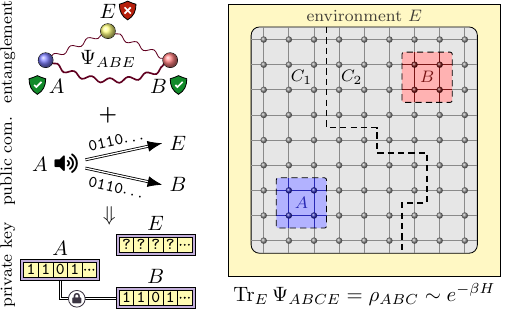}
    \caption{Left: In conventional few-body QKD protocols, two honest parties who have access to a bipartite entangled state $\rho_{AB}$ and a public classical communication channel distill perfectly correlated random classical bits (a \textit{key}) that are private from any eavesdropper $E$ holding degrees of freedom such that $\Psi_{ABE}$ is pure. Right: In the many-body setting, we instead consider QKD protocols that access two subregions $A$ and $B$ of a larger state $\rho_{ABC}$, with the eavesdropper holding degrees of freedom $E$ such that $\Psi_{ABCE}$ is pure. Successful distillation implies the presence of entanglement across all bipartitions $AC_1:BC_2$ that separate $A$ from $B$, where $C = C_1C_2$.}
    \label{fig:qkd}
\end{figure}

Entangled states are useful in cryptography because entanglement is monogamous: the more entangled two degrees of freedom are, the less correlated they can be with any other parties who may be eavesdropping \cite{nielsen2010quantum}. For instance, two parties who share Einstein-Podolsky-Rosen (EPR) pairs can perform local measurements to generate a \textit{private key} \cite{ekert1991quantum}, a uniformly distributed random string of bits known only to the observers who perform the measurements, which can be used to encrypt messages. More generally, keys can be extracted from the private correlations contained within generic, noisy entangled states. This process, known as QKD, involves local operations as well as a `public' communication channel to which eavesdroppers have access (see Fig.~\ref{fig:qkd}).

By considering QKD protocols, we introduce a framework to quantify the privacy of correlations in many-body quantum systems that are in thermal equilibrium with an environment. When privacy is defined relative to an eavesdropper having access to the environment degrees of freedom, the possibility of QKD implies that the system is entangled \cite{curty2004,haseler2008}. One of our central contributions is to relate the privacy of correlations, and hence entanglement, to standard probes of a thermal state's linear response \cite{kubo1966fluctuation}. Using this relation we show that the entanglement in many-body thermal states has clear signatures in the correlations between local observables at low finite temperatures.

Our findings complement the recent result that, in thermal states at sufficiently high temperatures, there is no entanglement even on the shortest length scales \cite{bakshi2024high}. The methods introduced here allow us to detect when entanglement emerges at various length scales as the temperature is lowered. Moreover, we reveal a strong dependence of entanglement on the presence of conserved charges: thermal states with fixed total charge (which describe the `canonical ensemble') are entangled at all finite temperatures, but even small fluctuations in the global charge can degrade this entanglement.

\textit{Setup.---}
We consider two spatially separated `honest parties' $A$ and $B$ who wish to communicate with one another while guaranteeing that no other party $E$ (the eavesdropper) can read their message.
They are given access to their respective parts of a state $\rho_{ABC}$, where $C$ is the complement of $AB$ in the system, while $E$ holds degrees of freedom such that the global state $\Psi_{ABCE} \equiv \ket{\Psi}\bra{\Psi}$ is pure. This state can be generated repeatedly, and we denote the total number of copies used by $R$. In addition, $A$ and $B$ can use a public classical communication channel, which $E$ can access but not modify. This setup is illustrated in Fig.~\ref{fig:qkd}.

Using these resources, the honest parties try to generate a shared private key. At large $R$, the key $\vec{k}= k_{1},\ldots,k_{KR}$ is uniformly random over all $2^{KR}$ possible $KR$-bit strings and is accessible to both $A$ and $B$; here $K$ is the \emph{key rate}, with $0 \leq K \leq 1$. The key is private from the eavesdropper if all the information held by $E$ at the end of the protocol (the quantum states and any publicly communicated data) is uncorrelated with $\vec{k}$ \footnote{For finite $R$, the key produced by a given QKD protocol will only be approximately private and uniform}.

\emph{QKD protocol.---} We focus on protocols in which parties $A$ and $B$ each perform local measurements on individual copies of $\rho_{ABE}$ and all public communication is `one way', i.e. from $A$ to both $B$ and $E$. The measurements performed by $A$ and $B$ in each of the $R$ rounds can be described by by positive operator-valued measures (POVMs) $M_{A,a}$ and $M_{B,b}$, respectively, which are local operators such that the joint distribution of outcomes $a$ and $b$ is $p_{ab}=\text{Tr}[M_{A,a}M_{B,b}\rho_{AB}]$. Afterwards, $A$ and $B$ each has a random string of $R$ bits, $\vec{a} = a_1, \ldots, a_{R}$ and $\vec{b} = b_1, \ldots, b_{R}$, while $E$ has access to $\rho_{E,a_1b_1} \otimes \cdots \otimes \rho_{E,a_Rb_R}$, where the density matrix $\rho_{E,ab} \propto \text{Tr}_{AB}[M_a M_b \rho_{ABE}]$.

In brief, QKD is possible if party $B$ can distill more information about $\vec{a}$ than party $E$ \cite{cai2004quantum,devetak2005distillation}. A standard result in classical Shannon theory implies that $B$ can extract $RI_{ab}$ bits' worth of information about $\vec{a}$ from their bitstring in the asymptotic limit $R \rightarrow \infty$, where $I_{ab} = H_a+H_b-H_{ab}$ is the mutual information, and e.g. $H_a = -\sum_a p_a\log_2 p_a$ is the Shannon entropy of the classical random variable $a$ \cite{Cover2005}.
Meanwhile, for any measurement strategy used by $E$, the mutual information between their outcomes and $\vec{a}$ can be no greater than $R\chi_E$, where
\begin{align}
    \chi_E = S(\rho_E) - \sum_a p_a S(\rho_{E,a})
    \label{eq:defchiE}
\end{align}
is the Holevo quantity \cite{nielsen2010quantum}; here $S(\rho)=-\text{Tr}[\rho \log_2 \rho]$ denotes the von Neumann entropy. Heuristically, if $I_{ab} > \chi_E$, then the discrepancy $(I_{ab} - \chi_E)R$ represents the amount of information contained in $\vec{a}$ that is available from $\vec{b}$, but inaccessible to $E$. 
 
Concretely, Refs.~\cite{cai2004quantum,devetak2005distillation} present a generic strategy (see \cite{SI} for an informal outline) for $A$ and $B$ to classically post-process $\vec{a}$ and $\vec{b}$ such that each ends up with a key $\vec{k}$ of length $KR$, where at large $R$ the key rate $K$ approaches
\begin{align}
    K = I_{ab} - \chi_E \leq \chi_B-\chi_E, \label{eq:keyrate}
\end{align}
where $\chi_B = S(\rho_B) - \sum_a p_a S(\rho_{B,a})$. We have included the inequality to highlight the fact that, if party $B$ is allowed to optimize their measurement strategy over POVMs acting on the state $\rho_{B,a_1} \otimes \cdots \otimes \rho_{B,a_R}$, instead of simply applying a fixed POVM in each round, it is possible to achieve a key rate $K$ equal to the `private information' $\chi_B-\chi_E$ \cite{nielsen2010quantum}. Note also that, although we are considering $R$-round protocols with $R$ large, the rate in Eq.~\eqref{eq:keyrate} is computed from properties of a single copy of the density matrix $\rho_{ABE}$.

As mentioned above, the success of a QKD protocol reveals entanglement. When the initial state of $E$ purifies $\rho_{ABC}$, a key rate $K>0$ implies that $\rho_{ABC}$ has bipartite entanglement across all bipartitions of $ABC$ which separate $A$ from $B$ \cite{curty2004,acin2005quantum,haseler2008,horodecki2009general,SI}; bipartitions of this kind are illustrated on the right of Fig.~\ref{fig:qkd}. An alternative interpretation is that, if there exists a measurement in $A$ which reduces the entropy of a subregion $B$ more than it reduces the entropy of the full state $ABC$ (i.e. $\chi_B > \chi_{ABC}=\chi_E$), then $\rho_{ABC}$ is entangled.

It is important to note that in the setup described above, where we distinguish $E$ and $C$, the key rate $K$ does not witness entanglement in the reduced density matrix $\rho_{AB}$. In fact, it is straightforward construct examples where $K>0$ because $\rho_{ABC}$ is entangled, but where $\rho_{AB}$ is exactly separable \cite{SI}.

\textit{Thermal states.---} We now discuss private correlations in thermal states. Denoting the Hamiltonian of $ABC$ by $H$, at temperature $T = 1/\beta$ the standard thermal density matrix of the system $\rho_{\beta} = Z_{\beta}^{-1} e^{-\beta H}$ where $Z_{\beta}=\text{Tr}e^{-\beta H}$. If $H$ is invariant under global unitary transformations $U$, it is also natural to consider the thermal ensemble within a symmetry sector. Identifying the symmetry sectors $q$ with corresponding projectors $\Pi_q$, i.e. $H = \sum_q \Pi_q H \Pi_q$, the thermal ensembles for individual sectors are $\rho_{\beta,q}=Z_{\beta,q}^{-1} e^{-\beta H}\Pi_q$ with $Z_{\beta,q}=\text{Tr} [e^{-\beta H}\Pi_q]$. While $\rho_{\beta}$ is weakly symmetric under $U$, i.e. $U \rho_{\beta} U^{\dag} = \rho_{\beta}$, $\rho_{\beta,q}$ are strongly symmetric: $U \rho_{\beta, q} = \rho_{\beta, q} U \propto \rho_{\beta, q}$. When the symmetry corresponds to $\mathrm{U}(1)$ charge conservation, $\rho_{\beta}$ is known as a grand canonical ensemble, while $\rho_{\beta,q}$ is a canonical ensemble. For a system density matrix $\rho$, which is here either $\rho_{\beta}$ or $\rho_{\beta,q}$, the pure state of $ABCE$ can be chosen as
\begin{align}
    \ket{\Psi} \propto [\sqrt{\rho} \otimes \mathbbm{1}_E] \ket{\Psi_{0}},
    \label{eq:CanonicalPurification}
\end{align}
where $\ket{\Psi_0}$ is a maximally entangled state between $ABC$ and $E$.

Notably, Ref.~\cite{bakshi2024high} recently showed that thermal states $\rho_{\beta}$ of locally interacting qubit systems are exactly separable for $\beta < \beta_s$ below a threshold $\beta_s$ that is finite in the thermodynamic limit. However, it is unclear whether this behavior has implications for standard observables. By relating private correlations in states of the form Eq.~\eqref{eq:CanonicalPurification} to finite temperature linear response, we will show in the following section that it does. In particular, we will see throughout this work that entanglement can be detected from two-point correlations  for $\beta > \beta_p$ where $\beta_p$ is finite in the thermodynamic limit, and $\beta_p > \beta_s$. 

\emph{Linear response.---}
By considering a scenario where party $A$ performs weak measurements of a local operator $\mathcal{O}_A$, and working at the lowest nontrivial order in the strength $\mu$ of these measurements, we will relate $\chi_E$ to the linear response of a thermal state $\rho_{\beta}$. A weak two-outcome measurement of $\mathcal{O}_A$ can be represented by the POVM
\begin{align}
    M_{A,a} = \frac{1}{2}\big( 1 + (-1)^a \mu \mathcal{O}_A\big), \label{eq:MAa}
\end{align}
where $a = 0,1$ and $\mu \ll 1$. The post-measurement states of $E$ are then $\rho_{E,a} = p_a^{-1} \rho^{1/2} M_s \rho^{1/2}$, where $p_a = \text{Tr}[\rho M_a]$. Expanding the expression for $\chi_E$ for such $M_{A,a}$ in powers of $\mu$, we find that $\chi_E = \frac{1}{2}\mu^2 \partial^2_{\mu} \chi_E + O(\mu^3)$ with \cite{SI}
\begin{align}
    [\ln 2]^{-1} \partial^2_{\mu} \chi_E = \frac{1}{2\pi} \int_{-\infty}^{\infty} d\omega \,  [\beta \omega b(\beta \omega)]  s_{\mathcal{O}_A}(\omega,\beta),
    \label{eq:chiEspectral}
\end{align}
where $s_{\mathcal{O}_A}(\omega,\beta) = \int_{-\infty}^{\infty}dt\, \braket{\mathcal{O}_A(t)\mathcal{O}_A}_{\beta,c} e^{i \omega t}$ is the spectral function of $\mathcal{O}_A$ at inverse temperature $\beta$, defined here as the Fourier transform of the connected autocorrelation function  \footnote{Thermal averages are denoted $\braket{\mathcal{O}}_{\beta}=\text{Tr}[\rho_{\beta}\mathcal{O}]$ while connected correlation functions between operators $\mathcal{O}$ and $\mathcal{O}'$ are defined by $\braket{\mathcal{O}\mathcal{O}'}_{\beta,c} \equiv \text{Tr}[\rho_{\beta}\mathcal{O}\mathcal{O}']-\text{Tr}[\rho_{\beta}\mathcal{O}]\text{Tr}[\rho_{\beta}\mathcal{O}']$}, and $b(\beta \omega) = [e^{\beta \omega}-1]^{-1}$ is the Bose function. To arrive at this expression we have used detailed balance $s_{\mathcal{O}_A}(-\omega,\beta)=s_{\mathcal{O}_A}(\omega,\beta)e^{-\beta\omega}$. Equation~\eqref{eq:chiEspectral} is one of our central results: for thermal states $\rho_{\beta} \sim e^{-\beta H}$ the information accessible to the eavesdropper is related to the relaxation of the observable $\mathcal{O}_A$. Since the state of $ABCE$ is pure, Eq.~\eqref{eq:chiEspectral} also describes the reduction in the von Neumann entropy of the thermal state $\rho_{\beta}$ by the measurement.

The classical mutual information $I_{ab}$ between outcomes of measurements in $A$ and $B$ depends on the measurement strategy. For the measurement of party $B$ it is convenient to consider the POVM $M_{B,b} = \frac{1}{2}( 1 + (-1)^b  \mathcal{O}_B)$ with $b=0,1$, where we have scaled $\mathcal{O}_B$ such that $||\mathcal{O}_B||_{\infty}=1$. At small $\mu$ we then find $I_{ab} = \frac{1}{2}\mu^2 \partial_{\mu}^2 I_{ab} + O(\mu^3)$, where as above the derivative is evaluated at $\mu=0$, and
\begin{align}
    [\ln 2]^{-1} \partial_{\mu}^2 I_{ab}  = \frac{ \braket{\mathcal{O}_A\mathcal{O}_B}_{\beta,c}^2}{1 - \braket{\mathcal{O}_B}_{\beta}^2},
    \label{eq:d2IAB}
\end{align}
where $\braket{\mathcal{O}_A\mathcal{O}_B}_{\beta,c}$ is a connected two-point correlation function in the thermal state $\rho_{\beta}$.

In ground states, and pure states more generally, any correlations $I_{ab} \neq 0$ between $\mathcal{O}_A$ and $\mathcal{O}_B$ imply entanglement across all bipartitions of $ABC$ separating $A$ from $B$. Our results in Eqs.~\eqref{eq:chiEspectral} and \eqref{eq:d2IAB} generalize this statement to finite temperatures: If $\partial_{\mu}^2 I_{ab} > \partial_{\mu}^2 \chi_E$ then $\rho_{ABC}$ is entangled across all bipartitions of $ABC$ separating $A$ from $B$. It is notable that entanglement in the global state $\rho_{ABC}$ can be detected via correlations and response functions of local observables in regions $A$ and $B$.

Remarkably, if the system is in a thermal state with respect to the Hamiltonian which generates its dynamics, Eqs.~(\ref{eq:chiEspectral}, \ref{eq:d2IAB}) provide a straightforward way to measure the lowest-order contribution to the key rate $\partial_\mu^2 K$ in experiment, and hence to detect bipartite entanglement.

\textit{Strong symmetry.---} We now discuss the effects of a strong symmetry on $\chi_E$. Consider a binary POVM of the form in Eq.~\eqref{eq:MAa} with arbitrary $\mu$. If there exists a unitary $U$ such that $U \rho \propto \rho$ and $U \mathcal{O}_A U^{\dag}=-\mathcal{O}_A$, from $\rho_{E,a}=p_a^{-1}\rho^{1/2}M_a \rho^{1/2}$ we see that $\rho_{E,0}=\rho_{E,1}$. Since $\rho_E = \sum_a p_a\rho_{E,a}$, from the definition of $\chi_E$ in Eq.~\eqref{eq:defchiE} we have $\chi_E=0$: In a strongly symmetric mixed state, correlations between symmetry-odd observables are private from $E$. Any $\braket{\mathcal{O}_A\mathcal{O}_B}_{\beta,c} \neq 0$ for odd $\mathcal{O}_A$ and $\mathcal{O}_B$ therefore implies entanglement across all bipartitions separating $A$ from $B$. This result can also be derived using the arguments in Refs.~\cite{lu2023mixed,chen2024symmetry}, see Ref.~\cite{SI}. 

Therefore, unlike $\rho_{\beta}$, strongly symmetric thermal states $\rho_{\beta,q}$ are entangled at all finite temperatures. This is the case whenever the Hamiltonian $H$ includes symmetry-even terms $\mathcal{O}_j\mathcal{O}_k+\cdots$ involving products of symmetry-odd operators $\mathcal{O}_j$ and $\mathcal{O}_k$ that act on different degrees of freedom ($j$ and $k$). For example, in a qubit system with $\mathrm{U}(1)$ symmetry generated by the charge operator $Q \coloneqq \sum_j Z_j$, $H$ can have contributions of the form $X_j X_k+Y_j Y_k$. If we choose our subregions $A$ and $B$ to be subregions $j$ and $k$, respectively, then writing $\mathcal{O}_A=\mathcal{O}_j$ and $\mathcal{O}_B=\mathcal{O}_k$ we have
\begin{align}
    \partial_{\mu}^2 I_{ab} \propto \beta^2 \text{Tr}[\rho_{0,q}\mathcal{O}^2_A\mathcal{O}^2_B]^2 + \cdots, \label{eq:dmuI}
\end{align}
where the ellipsis denotes other terms of order $\beta^2$ as well as terms higher order in $\beta$. Since $\chi_E=0$ for symmetry-odd $\mathcal{O}_A$ we have $K = O(\mu^2\beta^2)$ at small $\mu$ and $\beta$, but any $K > 0$ implies that $\rho_{ABC}$ is nonseparable across bipartitions separating $A$ and $B$. As an example, in the case of $\mathrm{U}(1)$ symmetry, this analysis shows that although $\beta_s$ and $\beta_p$ are finite in the grand canonical ensemble, for the canonical ensemble we have $\beta_s=\beta_p=0$.

We can gain a clearer picture of the interplay between conserved charges and entanglement by re-introducing small charge fluctuations. 
In the case of $\mathrm{U}(1)$ symmetry, $\rho_{\beta}$ exhibits fluctuations of the charge $Q = \sum_j Z_j$ having standard deviation $\Delta Q \sim N^{1/2}$, whereas  $\rho_{\beta,q}$ has $\Delta Q=0$ by definition. As an intermediate between these extremes, consider the state $\rho = (1-\epsilon)\rho_{\beta,q} + \epsilon \rho_{\beta,q+1}$, where the subscripts $q$ and $q+1$ denote eigenvalues with respect to $\sum_j Z_j$, which has $\Delta Q = \sqrt{\epsilon(1-\epsilon)} \ll N^{1/2}$. Choosing $\mathcal{O}_A=X_j$, and generalizing Eq.~\eqref{eq:chiEspectral} to this setting, we find $\partial_{\mu}^2 \chi_E \sim \epsilon \ln(1/\epsilon)$ at small $\epsilon$ and zeroth order in $\beta$. Setting $\mathcal{O}_B=X_k$ in the analog of Eq.~\eqref{eq:dmuI}, we then find that the condition $\partial_{\mu}^2 K>0$ is satisfied for $\beta > \beta_p$, with $\beta_p \sim [\epsilon \ln(1/\epsilon)]^{1/2}$. When the fluctuations of $Q$ are only of order unity, parametrically smaller than in $\rho_{\beta}$, there are no private correlations above a high but finite temperature. Although there is entanglement in high temperature states with strong symmetry, our analysis suggests that the entanglement is fragile to the breaking of the strong symmetry.

\textit{Low temperatures.---}We now discuss the behavior of $\partial_{\mu}^2\chi_E$ and $\partial_{\mu}^2 I_{ab}$ for $\rho \propto e^{-\beta H}$ when $H$ has a disordered, quantum critical, or long-range ordered ground state. Universal behavior is anticipated when the correlation length $\xi$ of the system is much larger than microscopic length scales, and so in the following we will focus on this regime. For concreteness, we consider a system which undergoes a continuous quantum phase transition from an ordered to a disordered ground state as a parameter $g$ is increased through a critical point $g_c$. 

At low temperatures on the disordered side ($g>g_c$), $\chi_E$ is controlled by the gap $\epsilon \sim (g-g_c)^{z\nu}$ to excitations. Here $\nu$ is the critical exponent characterizing the divergence of the $T=0$ correlation length $\xi \sim (g-g_c)^{-\nu}$ and $z$ is the dynamic critical exponent, with $z \geq 1$ in locally interacting systems \cite{sachdev2011quantum}. At $T \ll \epsilon$ the contributions to $s_{\mathcal{O}_A}(\omega,\beta)$ for $0 < \omega < \epsilon$ are associated with transitions from the first excited state to higher energy states, and these contributions are suppressed by the Boltzmann weight $\sim e^{-\beta \epsilon}$ in $s_{\mathcal{O}_A}(\omega,\beta)$. Contributions from transitions across the gap, for which $s_{\mathcal{O}_A}(\omega,\beta)$ may be of order unity, are meanwhile suppressed by $b(\epsilon) \approx e^{-\beta \epsilon}$. Therefore, Eqs.~\eqref{eq:chiEspectral} and \eqref{eq:d2IAB} give
\begin{align}
    \partial_{\mu}^2\chi_{E,\text{dis.}} \sim e^{-\beta\epsilon}, \quad \partial_{\mu}^2 I_{ab,\text{dis.}} \sim e^{-2x/\xi}, \label{eq:dis}
\end{align}
at low temperatures and large $x$, where $x$ is the separation between $A$ and $B$. Quite generally, a comparison of two-point spatial and temporal correlations is sufficient to detect entanglement (via the condition $\partial_{\mu}^2 K > 0$) at low finite temperatures. Here the length scale $x_K$ out to which QKD is possible, which we refer to as the key length, diverges with decreasing temperature as $x_K \sim \beta (g-g_c)^{(z-1)\nu}$ at large $\beta$.

\begin{figure}
    \includegraphics[width=0.47\textwidth]{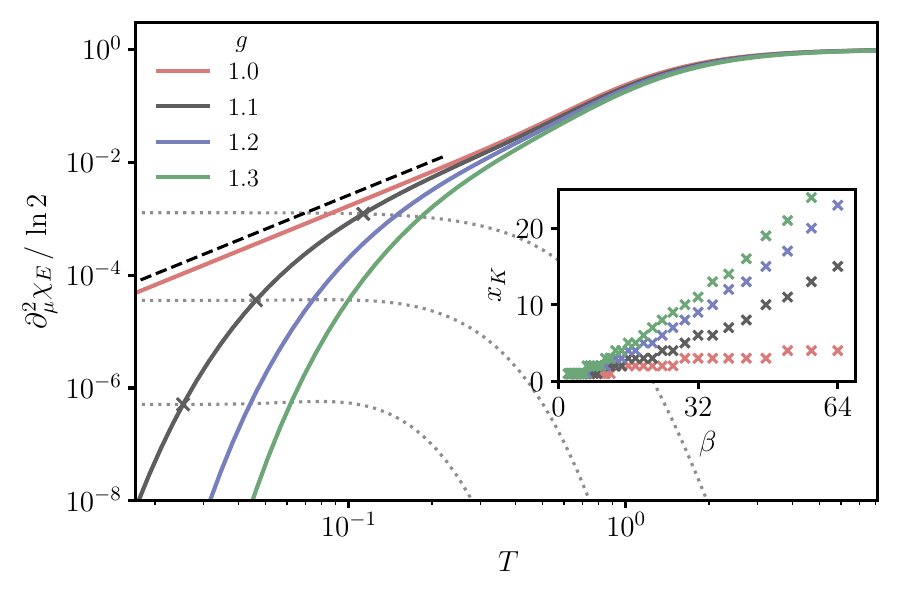}
    \caption{QKD from thermal states of infinite transverse field Ising chains. The Hamiltonian $H=-\sum_j(Z_j Z_{j+1}+g X_j)$, with $g=1$ a quantum critical point, $\mathcal{O}_A=X_0$, and $\mathcal{O}_B=X_x$ with $x \neq 0$. Solid lines in the main panel show the behavior of $\partial_{\mu}^2 \chi_E / \ln 2$ for various $g$ (legend) and dotted lines show $\partial_{\mu}^2 I_{ab} /\ln 2$ for $g=1.1$ and $x=2,4,8$ (top to bottom). The black dashed line shows the scaling $\partial_{\mu}^2 \chi_E \sim T^2$ expected at $g=1$. The inset shows the key length $x_K$, the largest $x$ for which $\partial_{\mu}^2 K > 0$, as a function of $\beta=1/T$.}
    \label{fig:Ising}
\end{figure}

If the ground state of $H$ spontaneously breaks a global symmetry, and for $\mathcal{O}_A$ and $\mathcal{O}_B$ local (scalar) order parameters, there is a zero frequency contribution to the spectral function $s_{\mathcal{O}_A}(\omega,\beta) = \mathcal{O}^2_{\text{ord.}}\delta(\omega)+\ldots$; here $\mathcal{O}_{\text{ord.}}$ is the ($\beta$-dependent) expectation value of the order parameter in an infinitesimal symmetry-breaking field, and the ellipsis denotes contributions from nonzero $\omega$. The implication is that, in ordered phases,
\begin{align}
    \partial_{\mu}^2\chi_{E,\text{ord.}} \approx \mathcal{O}^2_{\text{ord.}},\quad \partial_{\mu}^2 I_{ab,\text{ord.}}\approx \mathcal{O}^4_{\text{ord.}}, \label{eq:ord}
\end{align}
where we write approximate equalities because we have omitted contributions which vanish as $T \to 0$ and $x \to \infty$, respectively. Recalling that $||\mathcal{O}_{B}||_{\infty}=1$, in general we have $\mathcal{O}^2_{\text{ord.}} < 1$. Therefore, at large $x$, the lowest order contribution to the key rate $\partial_{\mu}^2 K = 0$ even as $T$ is decreased to zero: in weakly symmetric ordered systems, long-distance correlations are not private from the environment. By contrast, if parties $A$ and $B$ measure the order parameter in a strongly symmetric system we have $\chi_E=0$, so long-range order implies finite $K$ at arbitrarily large $x$. This implies that private correlations are sensitive to the entanglement reported in Ref.~\cite{lu2023mixed,chen2024symmetry}.

For systems with quantum critical ground states we can infer the behavior of $\partial_{\mu}^2\chi_E$ by dimensional analysis. For $\mathcal{O}_A$ which relaxes as $\braket{\mathcal{O}_A(t)\mathcal{O}_A}_{\beta,c} \sim t^{-2\Delta/z}$ for times $t \ll \beta$, with faster decay beyond this time \cite{sachdev2011quantum}, the integral in Eq.~\eqref{eq:chiEspectral} scales as $\beta^{-2\Delta/z}$ at low temperatures. If $\mathcal{O}_B$ is related to the local operator $\mathcal{O}_A$ by translation, then quite generally the connected correlations between the two operators decay as $\sim x^{-2\Delta}$ for $x \ll \xi$, where the finite-temperature correlation length $\xi \sim \beta^{1/z}$. Comparing $\partial_{\mu}^2\chi_E \sim \beta^{-2\Delta/z}$ with $\partial_{\mu}^2 I_{ab}$ reveals that, at low temperatures, the key length $x_K \sim \beta^{1/(2z)} \ll \xi$. When both $x$ and $x_K$ are much smaller than $\xi$, we therefore have
\begin{align}
    \partial_{\mu}^2\chi_{E,\text{crit.}} \sim \beta^{-2\Delta/z}, \quad \partial_{\mu}^2 I_{ab,\text{crit.}} \sim x^{-4\Delta}.\label{eq:crit}
\end{align}
The behavior outlined above is surprising: at finite temperatures, the power law correlations characteristic of the quantum critical system persist out to a distance $\sim \beta^{1/z}$, but at low temperatures these correlations between local observables can only be used to detect entanglement up to a parametrically smaller distance $\sim \beta^{1/(2z)}$. In fact, for quantum critical Hamiltonians whose large scale structure is described by conformal field theory, private correlations do survive out to a distance $\sim \xi$, but these correlations are hidden in nonlocal observables \cite{toappear}. 

\emph{Ising chain.---} In Fig.~\ref{fig:Ising} we calculate $\partial_{\mu}^2 \chi_E$ and $\partial_{\mu}^2 I_{ab}$ using exact results \cite{niemeijer1967exact} for the one-dimensional transverse field Ising model, having Hamiltonian $H = -\sum_j( Z_j Z_{j+1} + g X_j)$ with $X_j$ and $Z_j$ Pauli operators acting on qubit $j$. The ground state of this model undergoes a quantum phase transition, with $z=1$, from a state with long-range to short-range $Z$ correlations as $g$ is increased through $g_c=1$ \cite{sachdev2011quantum}. Our focus will be on privacy in the correlations between spatially separated transverse field operators, i.e. $\mathcal{O}_A = X_0$ and $\mathcal{O}_B = X_x$.

The main panel of Fig.~\ref{fig:Ising} shows the increase of $\partial_{\mu}^2\chi_E$ with $T$ for $g \geq g_c$; results for $g < g_c$ are similar. At $g=g_c$ and $T=0$, two-point spatial and temporal correlations between transverse field operators decay with exponent $2\Delta=2$, and at small $T$ the exact result in Fig.~\ref{fig:Ising} is in good agreement with the behavior $\partial_{\mu}^2 \chi_E \sim T^{2}$ predicted in Eq.~\eqref{eq:crit}. As expected, for $g > g_c$ we see a sharp decrease of $\partial_{\mu}^2 \chi_E$ as the temperature is decreased. In the inset of Fig.~\ref{fig:Ising} we calculate $x_K$ from a comparison between $\partial_{\mu}^2\chi_E$ and $\partial_{\mu}^2 I_{ab}$. As expected, at $g=g_c$ the large low-frequency spectral weight causes $x_K$ to increase more slowly with $\beta$ than for $g > g_c$. 

\textit{Entanglement distillation.---} We now compare QKD with entanglement distillation \cite{bennett1996mixed,Deutsch1996}, which is a strictly more demanding task: if EPR pairs can be distilled, one can always generate private keys by measuring these EPR pairs in a fixed basis \cite{devetak2005distillation}. An upper bound on the rate of entanglement distillation from many copies of $\rho_{AB}$ is provided by the logarithmic negativity $N_{AB} \coloneqq \log ||\rho_{AB}^{T_A}||_1$ \cite{vidal2002computable,plenio2005logarithmic}; here $T_A$ denotes a partial transpose on $A$, and $\|X\|_1 \coloneqq \Tr \sqrt{X^\dagger X}$. However, even in the ground states of physical many-body systems, the negativity between small subregions $A$ and $B$ generally vanishes when their separation $x$ exceeds a finite value \cite{javanmard2018sharp,parez2024entanglement}. Heuristically, this is because entanglement distillation requires that degrees of freedom in $A$ and $B$ are decoupled from both $C$ and $E$. In physical many-body systems, where correlations generally decay with separation, the strong correlations between $A$ (or $B$) and its immediate surroundings in $C$ mean that the degree of decoupling necessary for entanglement distillation cannot be achieved using simple local operations. In contrast, for QKD, we require privacy from $E$ but not $C$, which is why the local measurement schemes considered here can succeed at large $x$ and small $T$. 

\emph{Discussion.---} In this work we have shown that if the correlations between two observables in a thermal state exceed a temperature-dependent threshold, the state is nonseparable across all spatial bipartitions which separate the two observables. The threshold is set by the system's linear response. To arrive at this relation between standard correlations and entanglement, we asked whether local observers can distill correlations that are private from the environment, which here plays the role of an eavesdropper in QKD. 

In contrast to practical, few-body QKD schemes \cite{scarani2009security} (including those using small thermal states \cite{Newton2019,Walton2021}), in our work the eavesdropper's state purifies a large many-body state to which $A$ and $B$ have only local access. This formulation of a cryptographic task is targeted at developing a theoretical understanding of the role of local correlations in the entanglement of physical systems. Studies of quantum games \cite{daniel2021quantum,bulchandani2023playing,hart2024playing} and of quantum metrology \cite{Hyllus2012,Toth2012,hauke2016measuring} have provided complementary operational approaches to this problem. In particular, studies of the quantum Fisher information have provided another relation between entanglement and linear response in thermal states \cite{hauke2016measuring,SI}.

Beyond standard Gibbs states, our approach has provided a new way to study entanglement within different thermal ensembles. For example, we have shown that canonical ensembles of $\mathrm{U}(1)$-symmetric systems are generically nonseparable at all $\beta \neq 0$, and that there is entanglement beyond a finite $\beta$ when $\mathrm{U}(1)$ charge fluctuations are of order unity. It would be interesting to explore the connection between entanglement and charge fluctuations more generally and, in particular, when away from thermal equilibrium. When applying our results to nonequilibrium mixed states, including those generated by tracing out degrees of freedom from a larger pure state, it is important to note that our formula for the Holevo information of the environment \eqref{eq:chiEspectral} still applies. The only modification is that the spectral properties of $\beta H$ are replaced by those of $-\log\rho_{ABC}$, which is in some settings known as the entanglement Hamiltonian.  

\emph{Acknowledgements.---} The authors are grateful to Ehud Altman, Vir Bulchandani, David Huse, Sarang Gopalakrishnan and Zack Weinstein for useful comments and discussions. S.J.G. is supported by the Gordon and Betty Moore Foundation. M.M. acknowledges support from Trinity College, Cambridge, and from the US National Science Foundation (NSF) Grant Number 2201516 under the Accelnet program of Office of International Science and Engineering (OISE).

%\bibliography{PKrefs2.bib}

\newpage
\onecolumngrid
\setcounter{secnumdepth}{2}

\onecolumngrid

This supplemental information is organized as follows. In Sec.~\ref{sec:Heisenberg} we discuss different notions of entanglement relevant to many-body quantum systems, and their relation to standard correlation functions. Then, in Sec.~\ref{sec:SIprotocol} we outline a quantum key distillation (QKD) protocol, and in Sec.~\ref{sec:SIentanglement} we provide a self-contained explanation of why a nonzero key rate implies bipartite entanglement. In Sec.~\ref{sec:SIchiE} we derive the lowest order (in measurement strength) contribution to the Holevo quantity for the eavesdropper $E$, and in Sec~\ref{sec:SIchiB} we present the analogous result for the mutual information between measurement outcomes performed by the honest parties $A$ and $B$. In Sec.~\ref{sec:SIcanonical} we then provide an alternative derivation of our result that canonical thermal ensembles for generic quantum Hamiltonians are entangled at all finite temperatures. Following this, in Sec.~\ref{sec:SIQFI}, we discuss the relation between our cryptographic probe of entanglement and well-known diagnostics based on the quantum Fisher information (QFI). Finally, in Sec.~\ref{sec:SIising} we provide technical details necessary for the numerical calculations in the main text. 

\section{Entanglement and correlations in a ground state}\label{sec:Heisenberg}
Here we review the relation between entanglement and two-point correlation functions in a paradigmatic model of quantum matter. Let us focus on the ground state of the one-dimensional antiferromagnetic spin-1/2 Heisenberg model $H = \sum_j(X_j X_{j+1} + Y_j Y_{j+1} + Z_j Z_{j+1})$. The $\mathrm{SU}(2)$ symmetry here is not important for the general arguments, but will simplify the analysis. We take subregions $A$ and $B$ to each consist of a single qubit, and these qubits are separated in space by distance $x$ (for two adjacent qubits, $x=1$). As in the main text, the region $C$ is the complement of $A$ and $B$ within the system.

First, in pure states, any nonzero connected correlations between $A$ and $B$ imply that $\rho_{ABC}$ is entangled. Moreover, in the ground state of $H$ above, the connected correlations between Pauli operators such as $Z_A$ and $Z_B$ (supported on $A$ and $B$, respectively) decay roughly as a power law at large $x$. The fact that the ground state is entangled is therefore evident in two-point correlations between $A$ and $B$ even when they are well-separated.  

By contrast, the reduced density matrix $\rho_{AB}$ becomes exactly separable beyond a finite value of $x$. To see this we use the fact that, for a state of two qubits, if the partial transpose $\rho_{AB}^{T_A}$ of $\rho_{AB}$ is completely positive, then $\rho_{AB}$ is separable \cite{horodecki1996separability}. For the setup of interest here, the weak $\mathrm{SU}(2)$ symmetry of $\rho_{AB}$ allows us to write 
\begin{align}
    \rho_{AB} = -\braket{Z_AZ_B} \ket{\phi}\bra{\phi} + \big(1+\braket{Z_AZ_B}\big)(\mathbbm{1}/4),
\end{align}
where $\ket{\phi}$ is the two-qubit $\mathrm{SU}(2)$ singlet and $\braket{Z_AZ_B}=\braket{Z_AZ_B}_c$ is the connected correlation function in $\rho_{AB}$. From this expression it can be verified that $\rho_{AB}^{T_A}$ is completely positive, and hence $\rho_{AB}$ is exactly separable, for $\braket{Z_AZ_B} > -1/3$. In the ground state of this specific model, $\rho_{AB}$ is therefore separable for $x > 1$ \cite{sato2005correlation}, i.e. $\rho_{AB}$ is separable unless $A$ and $B$ are neighboring qubits. Since connected correlations typically decay with separation in many-body systems, two-qubit reduced density matrices are typically separable for $x$ beyond a finite threshold. 

The above discussion highlights the fact that, in pure states, there are two very different kinds of entanglement (bipartite entanglement in $\rho_{ABC}$ versus $\rho_{AB}$) having signatures in the correlations between observables local to $A$ and $B$. At finite temperatures (and in mixed states more generally) it is straightforward to study entanglement in the reduced density matrix $\rho_{AB}$ using the entanglement negativity, which is minus the sum of negative eigenvalues of $\rho_{AB}^{T_A}$. On the other hand, it is \emph{a priori} unclear how to generalize the statement, applicable to pure states, that connected correlations between $A$ and $B$ imply bipartite entanglement in $\rho_{ABC}$. As we have discussed in the main text, studies of private classical correlations provide us with this generalization. A nonvanishing key rate implies that $\rho_{ABC}$ is nonseparable across all bipartitions of $ABC$ separating $A$ from $B$. We discuss this further in Sec.~\ref{sec:SIentanglement}.

\section{Quantum key distillation protocol}\label{sec:SIprotocol}

Here we outline a generic QKD protocol that achieves the key rate $K = \chi_B - \chi_E$ quoted in the main text. This protocol was introduced and analyzed in full detail in Refs.~\cite{cai2004quantum,devetak2005distillation}, but here we will aim to provide a simple picture of its structure.

As explained in the main text, the scenario we consider features two observers, $A$ and $B$, each of who have access to non-overlapping (but not necessarily complementary) subregions of a thermal state $\rho=\rho_{ABC}$, and an eavesdropper $E$, who has access to some environment degrees of freedom which purify the entire state $\rho$. The QKD protocol consumes $R$ copies of the state and generates a $KR$-bit private key $\vec{k}$ for each of $A$ and $B$. Each $\vec{k}$ is generated with probability $\approx 2^{-KR}$. The protocol is designed such that the final state held by the eavesdropper, a tensor product over $R$ density matrices, is uncorrelated with $\vec{k}$.

The first step of the protocol is for $A$ to perform a measurement on each copy of the state. For simplicity, we restrict ourselves to binary measurements, which can be described by a two-outcome POVM $\{M_{0}, M_1\}$, with $M_1 = \mathbbm{1} - M_0$. Writing $a_r = 0,1$ for the measurement outcome on the $r^{\text{th}}$ copy, the probability that $A$ observes a string of measurement outcomes $\vec{a} = (a_1, \ldots, a_R)$ is $p_{\vec{a}} = \prod_{r=1}^R\Tr[M_{a_r}\rho_A]$, and the corresponding conditional states of $B$ and $E$ are, respectively,
\begin{align}
    \rho_{B,\vec{a}} \equiv \bigotimes_{r=1}^R \rho_{B,a_r}, \quad \quad \quad \rho_{E,\vec{a}} \equiv \bigotimes_{r=1}^R \rho_{E,a_r}
\end{align}
In the subsequent steps, $A$ and $B$ use a \textit{codebook} $\{c\}$ to extract their copies of the key $\vec{k}$; this codebook is known in advance to all parties. The codebook is a system for categorizing the $2^R$ possible bitstrings $\vec{a}$ into bins referred to as codes $c$. Each code $c$ in the codebook is a set of $\sim 2^{KR}$ distinct bitstrings, and we say that the bitstring $\vec{a}$ is a codeword of $c$ if $\vec{a} \in c$. To each codeword $\vec{a}$ of $c$ we associate a distinct key $\vec{k} = \vec{k}(c,\vec{a})$, a bitstring of length $KR$. This means that any party who knows both the code and the codeword can infer $\vec{k}$. 

Upon finding the measurement outcomes $\vec{a}$, the honest party $A$ randomly selects a code $c$ of which $\vec{a}$ is a codeword. They then publicly communicate the choice $c$ of code, but do not declare $\vec{a}$. Party $B$ then chooses a $c$-dependent measurement scheme such that they find outcome $\vec{k}$ with high probability. Meanwhile, the initial correlations in $\rho$ between subregions $A$ and $E$ are such that, when party $E$ knows $c$, there is no measurement scheme that they can choose which generates $\vec{k}$ with high probability (at large $R$). Since the maximum mutual information between $\vec{a}$ and the outcomes of measurements performed by party $B$ ($E$) is the Holevo quantity $\chi_B$ ($\chi_E$), it is natural to expect that $K=0$ unless $\chi_B > \chi_E$.

A technical result of Refs.~\cite{cai2004quantum,devetak2005distillation} shows that for any $K < \chi_B - \chi_E$ there always exists a codebook such that, with high probability, a code $c$ will be announced satisfying the following properties
\begin{enumerate}
    \item The conditional probabilities of the codewords $p(\vec{a}|c)$ are approximately uniform (\textit{evenness}).
    \item For all codewords $\vec{a} \in c$, the corresponding conditional states on $B$ are near-perfectly distinguishable, i.e.~there exists a $c$-dependent measurement process on $B$, described by a POVM with $\sim 2^{KR}$ elements $\Pi_{\vec{k}'}^{(c)}$ indexed by $\vec{k}'$, such that $\Tr[\Pi^{(c)}_{\vec{k}'} \rho_{B,\vec{a}}] \approx \delta_{\vec{k}'\vec{k}}$, where $\vec{k}=\vec{k}(c,\vec{a})$
    (\textit{goodness}).
    \item For all $\vec{a} \in c$, the corresponding conditional states on $E$ are approximately indistinguishable, i.e. knowledge of $c$ alone is not enough for party $E$ to determine $\vec{a}$ and hence $\vec{k}$: ${\rho_{E,\vec{a}} \approx \sum_{\vec{a} \in c}p(\vec{a}|c)\rho_{E,\vec{a}}}$ (\textit{secrecy})
\end{enumerate}
If these conditions are met, then at the end, all three parties will know $c$, but only the honest parties $A$ and $B$ have the means to learn which codeword of $c$ occurred. Only with this information can $A$ and $B$ compute $\vec{k}$, which is therefore private from $E$.

Note that in the above arguments, we have been imprecise with the degree of accuracy with which the above conditions must be met, as well as the probability with which these events must happen. We refer interested readers to Ref.~\cite{devetak2005distillation}, but roughly speaking, the protocol can achieve arbitrary degrees of accuracy and certainty for any constant $K < \chi_B - \chi_E$, as long as $R$ is taken to be correspondingly large. This demonstrates that the key rate $K=\chi_B-\chi_E$ is \textit{asymptotically} achievable. Note that even though this protocol involves many copies of $\rho$, its asymptotic rate can be calculated using the single copy (or `single-letter') quantities $\chi_B$ and $\chi_E$.

Achieving the rate $\chi_B-\chi_E$ generally requires party $B$ to optimize their measurement strategy over all operations on $\rho_{B,\vec{a}}$, a tensor product over $R$ density matrices for subregion $B$. To relate the key rate to standard correlations, and to simplify this problem, we can prescribe a (suboptimal) single-copy measurement strategy for party $B$. In that case, $\chi_B$ is replaced by the classical mutual information between the outcomes of measurements performed by the honest parties $A$ and $B$. To see this in a simple example, suppose that party $B$ applies a two-outcome POVM having elements ${M_b = \frac{1}{2}(1+(-1)^b\mathcal{O}_B)}$ in each round and only records the outcome $b$. The corresponding channel acts as $\rho_{B,a} \to \sum_{b=0,1} \text{Tr}[M_b \rho_{B,a}] \ket{b}\bra{b}$. After acting with this channel, the Holevo quantity $\chi_B$ for the state held by party $B$ is exactly $I_{ab}=H_a+H_b-H_{ab}$ where e.g. $H_{ab} = -\sum_{ab}p_{ab}\log_2 p_{ab}$ is the Shannon entropy for the probability distribution $p_{ab}=\text{Tr}[M_a M_b \rho]$. Within such a scheme, we can therefore achieve a key rate $K = I_{ab}-\chi_E$. Note that, because $\chi_B$ is the maximum of the classical mutual information optimized over all measurement strategies, $I_{ab} \leq \chi_B$.

\section{Nonzero key rate implies entanglement}\label{sec:SIentanglement}

In this section we present a self-contained derivation of the fact that $K=\chi_B-\chi_E\leq 0$ for separable states. This is a simple application of the quantum data processing inequality \cite{nielsen2010quantum}. A state of the system that is separable across the bipartition $A\bar{A}$, with $\mathcal{H}= \mathcal{H}_A \otimes \mathcal{H}_{\bar{A}}$, has the form
\begin{align}
	\rho = \sum_{i} p_{i} \ket{i_A}\bra{i_A} \otimes \ket{i_{\bar{A}}}\bra{i_{\bar{A}}},
\end{align}
with $p_{i} \geq 0$ and e.g. $\braket{i_A|i_A'} \neq \delta_{i i'}$ in general. Here $\bar{A} = B \cup C$, where $B$ is the subregion accessed by $B$, and $C$ is the complement of $A$ and $B$ within the system. A purification of this state, on $A\bar{A}E$ is $\ket{\Phi} = \sum_{i} p_{i}^{1/2}\ket{i_A}\ket{i_{\bar{A}}}\ket{e_{i}}$. Here $\ket{e_{i}}$ are a set of orthogonal states on $E$: $\braket{e_{i}|e_{i'}} = \delta_{i i'}$. 

When party $A$ performs their measurements and records the outcome $a$, the resulting classical-quantum-quantum state is
\begin{align}
	\sum_a \ket{a}\bra{a}\otimes \sqrt{M_a}\ket{\Phi}\bra{\Phi}\sqrt{M_a} = \sum_a \ket{a}\bra{a} \otimes \sum_{i i'} \big(p_{i} p_{i'}\big)^{1/2} \braket{i_A'|M_a|i_A} \ket{i_{\bar{A}}}\bra{i_{\bar{A}}'} \otimes \ket{e_{i}}\bra{e_{i'}},\label{eq:rhoRbarAE}
\end{align}
where $M_a$ are the elements of the (Hermitian) POVM. The Holevo quantities $\chi_B$ and $\chi_E$ are computed from this state;  note that the tensor product on the right-hand side is here ordered as $A\bar{A}E$.

Since $\rho$ is known by all parties in advance, the eavesdropper can adopt the following strategy. Given that the state has the form in Eq.~\eqref{eq:rhoRbarAE}, party $E$ performs a measurement in the orthogonal basis $\ket{e_i}$. If they find the outcome indexed by $i$, they prepare $\ket{i_{\bar{A}}}$. Following this, the density matrices of $A\bar{A}$ and $AE$ are identical: party $E$ has copied the conditional states of $\bar{A}$. Since the Holevo quantity for party $E$ cannot increase under the action of a channel by the data processing inequality, we therefore have $\chi_{\bar{A}} \leq \chi_E$.

Finally, observe that since $B \subseteq \bar{A}$, the density matrix accessible to party $B$ is obtained by `tracing out' subregion $C$. This operation is itself a quantum channel: the data processing inequality implies $\chi_B \leq \chi_{\bar{A}}$. Combining these inequalities we have
\begin{align}
    K = \chi_B - \chi_E \leq \chi_{\bar{A}} - \chi_{\bar{A}} = 0.
\end{align}
Conversely, if $\chi_B > \chi_E$, the state $\rho$ must not be separable across a bipartition separating $A$ from $B$.

\section{Holevo information for party $E$}\label{sec:SIchiE}

\newcommand{\dif}{{\rm d}}

Here we determine the leading-order behavior of the Holevo information for $E$ in the case of a weak measurement of strength $\mu$ in subregion $A$. As in the main text, we write the POVM for this measurement as $M_a = \frac{1}{2}(I+(-1)^a \mu \mathcal{O}_A)$, with $a = 0, 1$. Because $\chi_E$ is non-negative and vanishes at $\mu = 0$, the leading contribution is second order in $\mu$. Accordingly, we aim to evaluate the second derivative $\partial_\mu^2 \chi_E$ at $\mu = 0$.

Since $E$ consists of degrees of freedom that purify the state $\rho$ of the system, the post-measurement density matrices of $E$ are $\rho_{E,a}=p_a^{-1}\sqrt{\rho}M_a\sqrt{\rho}$ with $p_a = \text{Tr}[M_a \rho]$. The Holevo quantity $\chi_E=S(\rho)-\sum_a p_a S(\rho_{E,a})$ depends only on the eigenvalues of $\rho$ and $\rho_{E,a}$, and it is clear that $\rho_{E,a}$ is isospectral with $p_a^{-1} M_a \rho$ (assuming $\rho$ is non-singular). Therefore,
\begin{align}
    [\ln 2]^{-1} \partial_\mu^2 \chi_E  &= \partial^2_{\mu} \sum_a p_a \text{Tr}[(p_a^{-1}M_a \rho)\ln (p_a^{-1} M_a \rho)] \\&= -\partial_{\mu}^2 \sum_a p_a \ln p_a + \partial_{\mu}^2 \sum_a \text{Tr}\big[ M_a \rho \ln \big( M_a \rho\big)],\notag
\end{align}
where a factor $[\ln 2]^{-1}$ appears on the left because we define entropies in units of bits. It is straightforward to verify that the first term on the right-hand side of the second line is $-\partial_{\mu}^2 \sum_a p_a \ln p_a|_{\mu = 0} =-\text{Tr}[\rho \mathcal{O}_A]^2$.

The second term can be differentiated with respect to $\mu$ once using the trace identity $\partial_\mu\Tr[f(X)] = \Tr[(\partial_\mu X)f'(X)]$ for a differentiable function $f$ and $\mu$-dependent matrix $X$. We find
\begin{align}
    \partial_{\mu} \text{Tr}\big[ M_a \rho \ln \big( M_a \rho\big)] = \frac{1}{2}(-1)^a \text{Tr}\Big[ \mathcal{O}_A \rho \Big( 1 + \ln \big( M_a \rho\big)\Big)\Big].
    \label{eq:HolevoEFirstDeriv}
\end{align}
Setting $\mu=0$ in this expression it can be verified that, as stated above, $\partial_{\mu}\chi_E\big|_{\mu=0}=0$. To differentiate \eqref{eq:HolevoEFirstDeriv} again, following Ref.~\cite{Lieb1973}, we use the integral representation of the natural logarithm of a matrix $X$,
\begin{align}
    \ln X = \int_0^\infty \dif z [(1+z)^{-1}\mathbbm{1} - (X+z\mathbbm{1})^{-1}]
\end{align}
combined with $\partial_{\mu} [X^{-1}] = -X^{-1}[\partial_{\mu} X]X^{-1}$. Differentiating, setting $\mu = 0$, and summing over $a$ we get
\begin{align}
    \partial^2_{\mu}\sum_a \text{Tr}\big[ M_a \rho \ln \big( M_a \rho\big)]\Big|_{\mu = 0} &= \int_0^{\infty} dz\text{Tr}\Big[ \mathcal{O}_A\rho(\rho+z\mathbbm{1})^{-1}\mathcal{O}_A \rho(\rho+z\mathbbm{1})^{-1} \Big].
\end{align}
Inserting the spectral decomposition of the initial density matrix $\rho = \sum_j \lambda_j \ket{j}\bra{j}$ we then find
\begin{align}
    [\ln 2]^{-1} \partial_\mu^2 \chi_E\big|_{\mu=0} = \sum_{jk}|\braket{j|\mathcal{O}_A|k}|^2 \lambda_j\lambda_k \frac{\ln \lambda_k-\ln \lambda_j}{\lambda_k-\lambda_j} - \text{Tr}[\rho \mathcal{O}_A]^2.
\end{align}
For $\lambda_j = \lambda_k$ the summand should be understood as its limit for $\lambda_j \to \lambda_k$, i.e. as $|\braket{j|\mathcal{O}_A|k}|^2\lambda_j$. Since the limit $\lambda_j \rightarrow 0$ is well-defined, this expression can also be applied to cases where $\rho$ is singular.

For thermal states we can relate $\chi_E$ to the system's dynamics. In that case, $\lambda_i = e^{-\beta E_i}/Z_{\beta}$ where the partition $Z_{\beta} = \sum_i e^{-\beta E_i}$. We define the spectral function $s_{\mathcal{O}_A}(\omega,\beta)$ for the observable $\mathcal{O}_A$ as the Fourier transform of the connected autocorrelation function $\Tr[\mathcal{O}_A(t) \mathcal{O}_A \rho] - \Tr[\mathcal{O}_A \rho]^2$:
\begin{align}
    s_{\mathcal{O}_A}(\omega,\beta) = 2\pi \sum_{ij}\frac{e^{-\beta E_i}}{Z_{\beta}}\delta(E_i - E_j + \omega)|\braket{i|\mathcal{O}_A|j}|^2 - 2\pi \delta(\omega) \Tr[\rho \mathcal{O}_A]^2,
\end{align}
where the dependence on $\beta$ is implicit. Using the spectral function to replace the sum over eigenstates with an integral over frequencies $\omega$, we finally obtain
\begin{align}
    [\ln 2]^{-1} \partial_\mu^2\chi_E|_{\mu = 0} &= \frac{1}{2\pi}\int_{-\infty}^\infty \dif \omega\,  \frac{\beta \omega}{e^{\beta \omega}-1} s_{\mathcal{O}_A}(\omega,\beta). \label{eq:chiEspectralSI}
\end{align}
From detailed balance, $s_{\mathcal{O}_A}(-\omega,\beta) = e^{-\beta \omega}s_{\mathcal{O}_A}(\omega,\beta)$, we see that the integrand is symmetric under $\omega \rightarrow -\omega$. Since $ABCE$ is pure we have $\chi_E=\chi_{ABC}$, so  Eq.~\eqref{eq:chiEspectralSI} relates the measurement-induced reduction in the von Neumann entropy of a thermal state to the dynamics generated by the corresponding Hamiltonian. 

\section{Classical mutual information between measurements in $A$ and $B$}\label{sec:SIchiB}

In this section we calculate the classical mutual information between binary measurements performed by $A$ and $B$. If $A$ applies a two-element POVM with elements $M_{a} = \frac{1}{2}(1+ (-1)^a \mu\mathcal{O}_A)$ and $B$ applies a two-element POVM $M_{b} = \frac{1}{2}( 1 + (-1)^b \mathcal{O}_B/ ||\mathcal{O}_B||_{\infty})$, the mutual information between the outcomes $a$ and $b$ is related to connected correlations between $\mathcal{O}_A$ and $\mathcal{O}_B$ in the initial thermal state $\rho$. Note that $M_{b}$ is a positive operator because we have rescaled by the spectral norm $||\mathcal{O}_B||_{\infty}$. The joint probability distribution of $a$ and $b$ is
\begin{align}
	p(s_A,s_B) = \frac{1}{4}\Big( 1 + b \braket{\mathcal{O}_B} + \mu a\big[ \braket{\mathcal{O}_A} + b \braket{\mathcal{O}_A \mathcal{O}_B}\big]\Big),
\end{align}
and the (classical) mutual information is defined by $I_{ab} = H_a + H_b - H_{ab}$ with e.g. $H_a = -\sum_{a} p_a \log_2 p_a$. It can be verified that $\partial_{\mu} I_{ab}\big|_{\mu=0}=0$ and 
\begin{align}
	[\ln 2]^{-1} \partial_{\mu}^2 I_{ab}\big|_{\mu=0} = \frac{\big(\braket{\mathcal{O}_A \mathcal{O}_B} - \braket{\mathcal{O}_A}\braket{\mathcal{O}_B}\big)^2}{||\mathcal{O}_B||_{\infty}^2 - \braket{\mathcal{O}_B}^2}.\label{eq:d2IABSI}
\end{align}
For example, if $a$ and $b$ are unbiased (but in general correlated) random bits, corresponding to $\braket{\mathcal{O}_A} = \braket{\mathcal{O}_B} = 0$, this formula reduces to $[\ln 2]^{-1}\partial_{\mu}^2 I_{ab}\big|_{\mu=0} =  \braket{\mathcal{O}_A \mathcal{O}_B}^2$.

\section{Canonical ensembles are entangled at high finite temperatures}\label{sec:SIcanonical}

In the main text we showed that, for mixed states with strong symmetries, correlations between symmetry-odd observables imply bipartite entanglement. For example, canonical thermal ensembles are generically entangled at small finite $\beta$. Here we provide a simple alternative derivation of this result, focusing on the case of $\text{U}(1)$ symmetry. The arguments here are in the same spirit as those in Ref.~\cite{lu2023mixed}, and do not generalize straightforwardly to systems with nonvanishing fluctuations of the symmetry charge.

Consider a qubit system and a density matrix $\rho$ with strong $\text{U}(1)$ symmetry with generator $\sum_j Z_j$, such that $e^{i\theta\sum_j Z_j}\rho=e^{im\theta}\rho$ for integer $m$. Our proof is by contradiction. Suppose that the state $\rho$ of $A\bar{A}$ is separable across $A$ and $\bar{A}$,
\begin{align}
    \rho = \sum_i p_i \ket{i_A}\bra{i_A} \otimes \ket{i_{\bar{A}}}\bra{i_{\bar{A}}}.
\end{align}
Defining $Z_A = \sum_{j \in A}Z_j$ and $Z_{\bar{A}} = \sum_{j \in \bar{A}}Z_j$, the strong symmetry implies $e^{i\theta Z_A}\ket{i_A}=e^{i\theta m_{A,i}}\ket{i_A}$ and $e^{i\theta Z_{\bar{A}}}\ket{i_{\bar{A}}}=e^{i\theta m_{\bar{A},i}}\ket{i_{\bar{A}}}$, where the integers $m_{A,i}$ and $m_{\bar{A},i}$ satisfy $m_{A,i}+m_{\bar{A},i}=m$. Now suppose that $\rho$ has nonvanishing correlations between symmetry odd operators, such as $X_j$ and $X_k$ with $j \in A$ and $k \in \bar{A}$, i.e. $\text{Tr}[\rho X_j X_k] \neq 0$. If the Hamiltonian has a contribution $X_j X_k + Y_j Y_k$, correlations $\text{Tr}[\rho X_j X_k] \neq 0$ appear at first order in the high temperature expansion of the thermal density matrix. However, assuming $\rho$ has the form indicated above, we have
\begin{align}
    \text{Tr}[\rho X_j X_k] = \sum_i p_i \braket{i_A|X_j|i_A} \braket{i_{\bar{A}}|X_j|i_{\bar{A}}} = 0, 
\end{align}
which follows from the fact that e.g. $e^{i\theta Z_A}\ket{i_A}=e^{i\theta m_{A,i}}\ket{i_A}$. Because $\ket{i_A}$ has well-defined eigenvalue with respect to $Z_A$, the matrix elements $\braket{i_A|X_j|i_A}=0$. Therefore, if $\rho$ has strong $\text{U}(1)$ symmetry and nonvanishing correlations of any odd operators across the bipartition $A\bar{A}$, it cannot be separable across that bipartition. The implication is that, without fine tuning, canonical ensembles are nonseparable at all finite temperatures.

\section{Quantum Fisher Information}\label{sec:SIQFI}
Here we discuss the relation between our probe of entanglement, which acquires operational meaning through cryptography, to the QFI $F$, which plays an important role in metrology and which can detect multipartite entanglement \cite{Hyllus2012,Toth2012}. Recall that, when working to lowest order in the measurement strength $\mu$, our condition for a key rate $K > 0$ can be stated as
\begin{align}
    \frac{\braket{\mathcal{O}_A \mathcal{O}_B}_c^2}{||\mathcal{O}_B||_{\infty}^2-\braket{\mathcal{O}_B^2}} > \int \frac{d\omega}{2\pi} \frac{\beta \omega}{e^{\beta \omega}-1} s_{\mathcal{O}_A}(\omega,\beta)\label{eq:SIcryptocondition}
\end{align}
in the case where public communication is from $A$ to $BE$. If the above inequality is satisfied, then the system density matrix $\rho_{ABC}$ is entangled across all bipartitions of the system separating $A$ from $B$. 

As discussed some time ago in Ref.~\cite{hauke2016measuring}, the QFI provides another connection between entanglement and linear response in thermal states. The standard application involves an operator that is a sum of local Pauli operators, and here we denote this operator by $\Sigma = \sum_i Z_i$. The QFI for a thermal state is then given by \cite{hauke2016measuring}
\begin{align}
    F = \frac{1}{4\pi}\int_{-\infty}^{\infty} d \omega e^{-\beta \omega} [f(\beta \omega)/b^2(\beta \omega)] s_{\Sigma}(\omega,\beta),
\end{align}
where $f(\beta \omega)=[e^{\beta \omega}+1]^{-1}$ is the Fermi function. Now let us define multipartite entanglement. A pure state of an $N$ qubit system is $k$-producible if it can be expressed as a tensor product over states of subsets of the qubits, with none of these subsets having size larger than $k$; if a state is $k$-producible but not $(k-1)$-producible, it is $k$-partite entangled. As shown in Refs.~\cite{Hyllus2012,Toth2012}, for a $k$-partite entangled state the QFI must satisfy
\begin{align}
    F \leq \Big\lfloor \frac{N}{k}\Big\rfloor k^2 + \Big( N - \Big\lfloor \frac{N}{k}\Big\rfloor k\Big)^2.
\end{align}
A violation of this inequality therefore implies $(k+1)$-partite entanglement. 

To compare these condition on the QFI with our cryptographic condition for bipartite entanglement in  Eq.~\eqref{eq:SIcryptocondition}, we consider a slightly different setting. Let us replace $\Sigma$ above by $\mathcal{O}_A+\mathcal{O}_B$, with each of $\mathcal{O}_A$ and $\mathcal{O}_B$ Pauli operators. The condition for bipartite entanglement based on the QFI is then
\begin{align}
    \frac{1}{4\pi}\int_{-\infty}^{\infty} d\omega e^{-\beta \omega}[f(\beta \omega)/b^2(\beta \omega)] s_{\mathcal{O}_A+\mathcal{O}_B}(\omega,\beta) > 2, \label{eq:QFIbipartite}
\end{align}
i.e. if this inequality is satisfied, $\rho_{ABC}$ is entangled across bipartitions separating $A$ from $B$. Note that Eq.~\eqref{eq:QFIbipartite} differs from Eq.~\eqref{eq:SIcryptocondition} even when the state is pure. 

\section{Exact calculations in the Ising chain}\label{sec:SIising}
Here we describe the calculations of $\partial_{\mu}^2\chi_E$ and $\partial_{\mu}^2 I_{ab}$ in infinite one-dimensional Ising chains $H = -\sum_j (Z_j Z_{j+1}+g X_j)$, with $\mathcal{O}_A=X_0$ and $\mathcal{O}_B=X_x$. These calculations are based on the exact results for two-point correlation functions of the transverse field operator in Ref.~\cite{niemeijer1967exact}. To determine $\partial_{\mu}^2\chi_E$ we use Eq.~\eqref{eq:chiEspectralSI} and
\begin{align}
	\braket{X_0(t) X_0}_{\beta,c} &= \Bigg( \int_0^{\pi} \frac{d\varphi}{\pi} \Big[ \cos[\Lambda(\varphi) t] - i \sin [\Lambda(\varphi) t] \tanh[\beta \Lambda(\varphi)/2]\Big] \Bigg)^2 \\
	&- \Bigg( \int_0^{\pi} \frac{d\varphi}{\pi}\cos[2\lambda(\varphi)] \Big[ i\sin[\Lambda(\varphi) t] - \cos [\Lambda(\varphi) t] \tanh[\beta \Lambda(\varphi)/2] \Bigg)^2,\notag
\end{align}	
where $\varphi$ is a momentum. The dispersion of excitations $\Lambda(\varphi) = 2\sqrt{ (\cos \varphi-g)^2 + \sin^2\varphi}$, and $\tan[2\lambda(\varphi)] = \sin \varphi / (\cos \varphi -g)$ with $0 \leq \lambda(\varphi) \leq \pi$. Fourier transforming the above expression we find the spectral function $s_A(\omega) = \int_{-\infty}^{\infty} dt e^{i \omega t} \braket{X_0(t) X_0}_{\beta,c}$ from which we can then calculate
\begin{align}
	[\ln 2]^{-1} \partial_{\mu}^2 \chi_E\big|_{\mu=0} = \frac{1}{2\pi} \int_0^{\pi} d\varphi \int_0^{\pi} d\varphi' \big[ 
	&B\big(\Lambda(\varphi)+ \Lambda(\varphi')\big) (1 -C(\varphi) C(\varphi')) ( 1 + D(\varphi))(1+D(\varphi')) \label{eq:chiEintegral}\\
	+&B\big(-\Lambda(\varphi)-\Lambda(\varphi')\big)(1 -C(\varphi) C(\varphi')) ( 1 - D(\varphi))(1-D(\varphi'))\notag\\
	+&2B\big(\Lambda(\varphi)-\Lambda(\varphi')\big)(1 + C(\varphi) C(\varphi')) ( 1 + D(\varphi))(1 - D(\varphi'))\big],\notag
\end{align}
where for brevity we have defined $B(\omega) = \beta \omega/(e^{\beta \omega}-1)$, $C(\varphi) = \cos[2\lambda(\varphi)]$ and $D(\varphi)=\tanh[\beta \Lambda(\varphi)/2]$. Our results for $\partial_{\mu}^2\chi_E$ are based on the numerical evaluation of the double integral on the right-hand side of Eq.~\eqref{eq:chiEintegral}. To determine $\partial_{\mu}^2 I_{ab}$ using Eq.~\eqref{eq:d2IABSI} we require $\braket{X_0}_{\beta}=\frac{1}{\pi}\int_0^{\pi} d\varphi\, C(\varphi)D(\varphi)$ and, defining  $S(\varphi)=\sin[2\lambda(\varphi)]$, 
\begin{align}
\braket{X_0 X_{x \neq 0}}_{\beta,c} = -\Bigg(\int_0^{2\pi} \frac{d\varphi}{2\pi}\, e^{i x \varphi} C(\varphi)D(\varphi)\Bigg)^2 -\Bigg(\int_0^{2\pi} \frac{d\varphi}{2\pi}\, e^{i x \varphi} S(\varphi) D(\varphi)\Bigg)^2.
\end{align}

\end{document}